\documentclass[english,journal=nalefd,layout=twocolumn,manuscript=letter]{achemso}
\usepackage[T1]{fontenc}
\usepackage[utf8]{inputenc}
\usepackage{color}
\usepackage{babel}
\usepackage{textcomp}
\usepackage{amsmath}
\usepackage{amssymb}
\usepackage{graphicx}
\usepackage{hyperref}

\usepackage{braket}

\newcommand{\eps}{\varepsilon}

\makeatletter

\title{Ligand Induced Size-Dependent Circular Dichroism in Quantum Dots}

\author{Daniel Chabeda}
\affiliation{Department of Chemistry, University of California, Berkeley, California
94720, United States}
\email{daniel_chabeda@berkeley.edu}

\author{Stephen Gee}
\affiliation{Department of Materials, University of California, Santa Barbara, CA 93106-5050}
\email{stephengee@ucsb.edu}

\author{Eran Rabani}
\affiliation{Department of Chemistry, University of California, Berkeley, California
94720, United States}
\altaffiliation{Materials Sciences Division, Lawrence Berkeley National Laboratory,
Berkeley, California 94720, United States}
\alsoaffiliation{The Sackler Center for Computational Molecular and Materials Science,
Tel Aviv University, Tel Aviv, Israel 69978}
\email{eran.rabani@berkeley.edu}

\setkeys{acs}{articletitle = true}
\usepackage{bm}
\AbstractOn
\SectionsOn

\makeatother

\begin{document}

\begin{abstract}
Recent experiments have probed the chiral properties of semiconductor nanocrystal (NC) quantum dots (QDs), but understanding the circular dichroism lineshape, excitonic features, and ligand induction mechanism remains a challenge. We propose an atomistic pseudopotential method to model chiral ligand passivated QDs, computing circular dichroism (CD) spectra for CdSe QDs (2.6 - 3.8 nm). We find strong agreement between calculated and measured line shapes, predicting consistent bisignate lineshapes with decreasing CD magnitude as size increases. Our analysis reveals the origin of bisignate lineshapes, arising from nondegenerate excitons with opposing angular momenta. We also explore the impact of chiral ligand orientation on QD surfaces, observing changes in optical activity magnitude and sign. This orientation sensitivity offers means to distinguish ordered from disordered ligand configurations, facilitating the study of order-disorder transitions at ligand-QD interfaces.\\
\end{abstract}

Semiconductor QDs have revolutionized electronic and photonic devices by enabling selective and tunable control of charge and light based on their size.\cite{alivisatos_perspectives_1996,efros_electronic_2000,liu_colloidal_2021, bukowski_quantum_2002} A current research frontier is developing sustainably manufactured, highly flexible QDs that exhibit tunable chiroptical properties through their size, shape, and structure, giving access to simultaneous control over charge, light, and spin.\cite{konstantatos_ultrasensitive_2006, hao_optically_2021,cheng_optically_2018, crassous_materials_2023,nozik_quantum_2002,coe_electroluminescence_2002} Such chiral control has been proposed as a tool for enhanced biological imaging, computer vision, and spin-based computing, serving as a motivation to understand the fundamental properties of chiral QDs.\cite{crassous_materials_2023,medintz_quantum_2005, xia_chiral_2011,baimuratov_quantum-dot_2013,economou_scalable_2012,weiss_coherent_2012,baimuratov_chiral_2016}

The CD spectra of QDs show unique features compared to molecular systems. First, the low energy CD features of chiral ligand passivated QDs appear at optical wavelengths, indicating induction of chirality from the UV-absorbing surface ligands into the excitonic transitions of the NC.\cite{han_optical_2024} Secondly, each excitonic CD signal vanishes at the absorption maximum wavelength and has a bisignate shape, an observation referred to as the "derivative" lineshape.\cite{ben-moshe_probing_2016,kuznetsova_ligand-induced_2021,tohgha_ligand_2013} Ben-Moshe {\it et al}. proposed that attaching chiral capping ligands to an otherwise achiral QD perturbs the nanocrystal electronic structure, splitting the absorbing exciton into two chiral sublevels, preferentially excited by opposite circular polarizations.\cite{ben-moshe_probing_2016}  No theoretical examination of this proposed chirality-induced degeneracy lifting has been offered in the literature to the best of our knowledge, mainly because calculating accurate excited states in nanoscale systems of experimentally relevant sizes remains a computational challenge. Dipole-coupling models have been used to theoretically predict the magnitude of optical activity for nanocuboids,\cite{baimuratov_dislocation-induced_2015,tepliakov_optical_2016, tepliakov_optical_2019, gao_distinct_2018} and a recent work utilized first-principle methods to compute chiroptical spectra of a 1.9 nm CdSe QD.\cite{han_optical_2024} Additional insights from theory are required to clarify the interplay of ligand structure and chiroptical properties in chiral QDs, and specifically, it is necessary to develop computationally efficient methods that are capable of capturing the atomistic structure that generates chiroptical response while scaling modestly with system size.

To address this, we use the atomistic semiempirical pseudopotential approach to calculate circular dichroism spectra of CdSe QDs of relevant sizes, achieving excellent agreement with experimental results. Our results indicate that the excitonic peaks are not split into angular momentum sublevels by the chiral ligand field. Instead, neighboring excitonic transitions with differently signed angular momenta produce transitions with opposing rotational strength, and the CD lineshape depends on the competing contributions from many transitions across the entire spectrum. Beyond investigating the excitonic origin of the derivative lineshape, we apply our molecular level control over individual ligands to investigate orientation dependence. We find that the CD lineshape is sensitive to the orientation of chiral surface ligands, exhibiting angle-dependent and anisotropic peak intensities, and show that incorporating increased ligand disorder in ensemble averaged conditions leads to cancellations that attenuate the CD signal by up to three orders of magnitude.

Fundamental understanding of ligand order is central to optimizing properties of QDs because favorable optoelectronic properties are typically achieved through surface passivation.\cite{ben_moshe_size_2011, ben-moshe_chirality_2013, ben-moshe_probing_2016, ben-moshe_chiral_2012, moloney_chiral_2007,tohgha_ligand_2013, hao_ligand-induced_2020,boles_surface_2016,brown_energy_2014,wtoolan_controlling_2020} Ligand shell ordering impacts the QD growth, solution-phase stability, optical blinking rates, and radiative lifetimes, making the study of ligand organization of great technological and fundamental interest.\cite{wuister_luminescence_2004,widmer-cooper_orientational_2014,peng_shape_2000, widmer-cooper_ligand-mediated_2016,calvin_role_2022, calvin_thermodynamics_2022,balan_unsaturated_2019}  However, probing the order-disorder transition experimentally has required intensive techniques such as sum-frequency generation,\cite{watson_probing_2019, morris-cohen_chemical_2013,frederick_surface-amplified_2011} X-ray photoelectron spectroscopy,\cite{frederick_surface-amplified_2011} powder X-ray diffraction,\cite{calvin_observation_2021} and NMR.\cite{fisher_measurement_2019, morris-cohen_chemical_2013}  These techniques enable the assessment of relative orientation, either through a signal that requires inversion symmetry breaking or by spatially resolving the 3D chemical environment. Optical activity is a phenomenon that is sensitive to inversion symmetry breaking of chiral systems and has the potential to be employed as a less intensive method to study collective ligand orientation on QDs.\cite{cai_enhancing_2023,allenmark_induced_2003} To our knowledge, no experimental analyses of ligand orientation and order through optical activity measurements have been reported. In this report, in addition to uncovering the relation between the exciton structure and the CD response, we also highlight the possibility of CD to probe \textit{in situ} ligand order.

\begin{figure*}[t]
    \centering
    \includegraphics[width=0.9\linewidth]{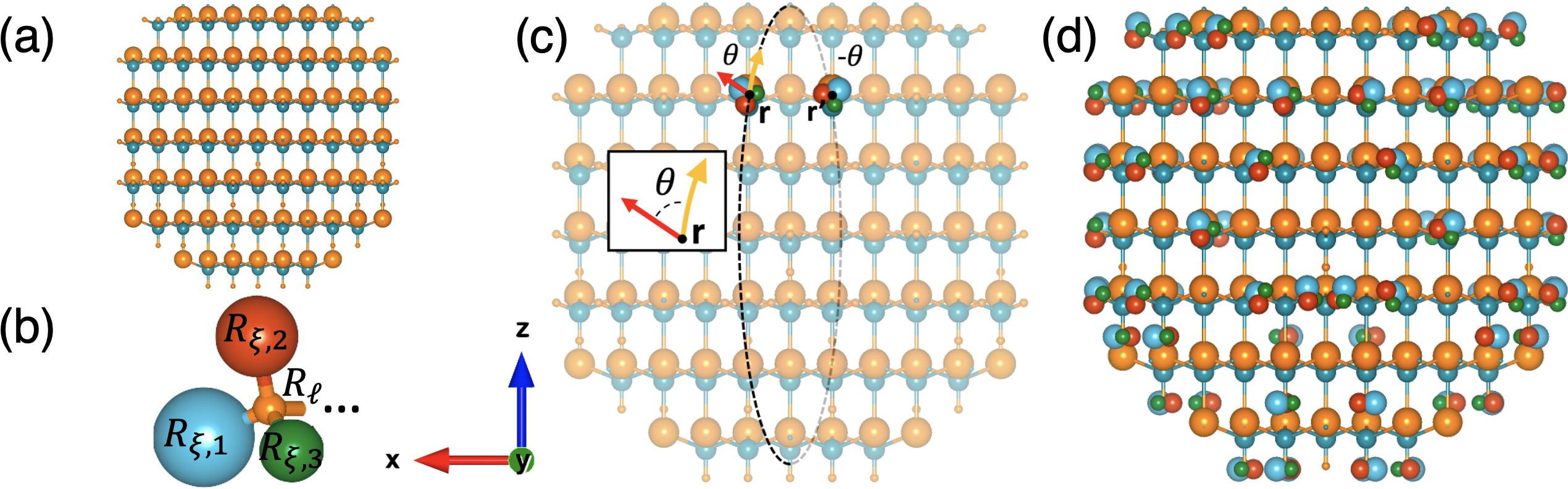}
    \caption{(a) Achiral 2.6 nm CdSe QD with our standard passivation ligand potentials (small orange and blue spheres) as viewed down the [$010$] axis. (b) The geometry of a CLP: the passivation potential attached to an undercoordinated Cd atom, $R_\ell$, becomes a stereocenter through the placement of three nonequivalent Gaussian potentials represented pictorally as blue, red, and green lobes. (c) Addition of a chiral ligand potential (CLP) with orientation angle $\theta$, where $\theta$ is the angle between a vector from the CLP center to its strongest lobe (red) and a geodesic toward the north pole (gold). To eliminate undesired structural asymmetry, we add a CLP with angle $-\theta$ across the two-fold symmetry plane, and all ligands are positioned in such pairs. (d) Mirror-symmetric CLPs are placed on the QD to generate the reported CD spectra.}
    \label{fig:passivation}
\end{figure*}

In order to describe the quasiparticle properties, we use the atomistic semi-empirical local pseudopotential model~\cite{jasrasaria_simulations_2022}, with a Hamiltonian, $\hat{h}_{QP}(\mathbf{r})$, given by:

\begin{align}
    \hat{h}_{QP}&(\mathbf{r}) = -\frac{1}{2} \nabla^2_{\mathbf{r}} + \sum_\alpha v_\alpha \left(|\mathbf{r - R}_{\alpha}| \right) \nonumber \\
    &+ 
    \sum_\ell  v_{\ell} \left(|\mathbf{r - R}_{\ell}| \right) + 
    \sum_\xi  v_{\xi} \left(\mathbf{r} \right) 
    \label{eq:quasiham}
\end{align}

\noindent where $\alpha$ indexes the atom centered at position $\mathbf{R}_{\alpha}$ with a pseudopotential $v_\alpha \left(|\mathbf{r - R}_{\alpha}\right)$, $\ell$ indexes the passivation ligand at position $\mathbf{R}_\ell$ with potential $v_\ell\left(|\mathbf{r - R}_{\ell}| \right)$, and $\xi$ indexes the chiral ligand potential (CLP) with potential $v_\xi \left(\mathbf{r} \right)$. The atom-centered pseudopotentials depend on the atom type, and for atom type $\mu$, it is given in reciprocal space by:\cite{wang_electronic_1999, wang_pseudopotential_1996} 
\begin{equation}
    \Tilde{v}_\mu(\mathbf{q}) = \frac{a_{0,\mu}(q^2 - a_{1,\mu})}{a_{2,\mu} \text{exp}(a_{3,\mu} q^2)-1}.
    \label{eq:kspacepseudopotl}
\end{equation}
In the above, $q$ is the reciprocal lattice vector and the $a_{i,\mu}$ are atom specific parameters (tabulated in Supporting Information) fitted to reproduce the bulk band structure, band gaps, effective masses, etc. We also use ligand pseudopotentials, $v_{\ell}(|\mathbf{r} - \mathbf{R}_{\ell}|)$, to passivate surface atoms with dangling bonds (Fig. \ref{fig:passivation}a).\cite{wang_electronic_1994} The procedure to place the ligand potential at position $\mathbf{R}_{\ell}$ is described elsewhere.\cite{jasrasaria_simulations_2022}

Since explicit description of the chiral ligand molecules is prohibited, to induce chirality, we add to the aforementioned passivation potentials a layer of chiral ligand pseudopotentials (CLPs). We take a simple form for these potential functions: a tetrahedral arrangement of three nonequivalent Gaussian functions (Fig.~\ref{fig:passivation}b)\cite{rabani_electronic_1999}
\begin{equation}
    v_{\xi}(\mathbf{r}) = \sum_{i=1,2,3} a_{i} \text{exp}\left(- b_i |\mathbf{r - R}_{\xi, i}|^2\right)
    \label{eq:chiralligpot}
\end{equation}
where $i$ indexes the three lobes of CLP $\xi$, each centered at $R_{\xi, i}$, needed to form a tetrahedron around passivation ligand $R_\ell$ (see Fig.~\ref{fig:passivation}b). The values of $a_{i}$ and $b$ are knobs that tune the magnitude and disymmetry of induced optical activity in our model (see Supporting Information Section S1). Small amplitudes $a_{i}$ were used such that the CLPs only weakly perturb the QD electronic structure, generating circular dichroism without significantly changing the band gap or quasiparticle energies (see Supporting Information).

We define the orientation of a CLP around the central ligand at position $\mathbf{R_\ell}$ by an angle, $\theta$, that is defined between the vector from $\mathbf{R_\ell}$ to the strongest lobe of the CLP (Fig.~\ref{fig:passivation}c red) and the vector along the geodesic from $\mathbf{R_\ell}$ toward the positive $z$~pole (Fig. \ref{fig:passivation}c gold). The choice of angle reference is arbitrary, but defining the angles with respect to a geodesic takes account of the natural spherical symmetry of the QDs. 

Introducing chiral ligands induces optical activity through the stereochemistry of the CLP. Additionally, because we perform calculations on a single QD rather than an ensemble, any other rotoreflection symmetry breaking will also induce optical activity. For example, an unintended "structural chirality" can arise from asymmetric placement of ligands across the $\sigma_2$ plane of the QD. This structural contribution is cancelled out over ensemble experimental measurements, but must be delineated and removed for calculations of individual QD particles. Computational works where this effect is not considered should carefully interpret the contributions from ligand stereochemistry vs structural symmetry breaking.\cite{han_optical_2024} 

To obtain a CD spectrum that captures ligand-induced chirality and excludes structural asymmetry, QD configurations must be generated in pairs that have structurally mirrored ligands, but consistent stereochemistry; ligands on QD A would have angle $\theta$ while the ligands on dot B would be oriented at $-\theta$ (Fig~\ref{fig:passivation}c). By enforcing the same stereochemistry for ligands on both dots, the ligand-induced chirality would persist while structural asymmetries are eliminated. An approximate workaround that avoids computing twice as many QD configurations and produces qualitatively similar results is to place structurally mirrored ligands on the same QD. We take this approach and populate half of the QD with CLPs at angle $\theta$, and complementary CLPs across the $\sigma_2$ plane at angle $-\theta$ (Fig. \ref{fig:passivation}d).

With our model Hamiltonian, we use a real-space grid representation and the filter-diagonalization technique~\cite{wall_extraction_1995,toledo_very_2002} to obtain valence and conduction band edge quasiparticle eigenstates with energy errors converged to within $10^{-3}$~meV. The excitation energy of each electron-hole pair were corrected to first order in the electron-hole interaction,\cite{brus_electronelectron_1984} and the optical absorption spectra was then calculated in the dipole approximation:
\begin{equation}
    \eps(\omega) = \frac{8 \pi \omega}{\hbar c}  \sum_{a,i} \delta(\omega - \omega_{ai}) |\pmb{\mu}_{ai}|^2,
    \label{eq:absorptionspectra}
\end{equation}
where $c$ is the speed of light, $\omega_{ai}$ is the correlated excitation energy of exciton $|a\rangle \otimes |i\rangle$ having an electronic transition dipole given by:
\begin{equation}
    {\boldsymbol{\mu}}_{ai} = {\langle a|\mathbf{r}|i} \rangle.
    \label{eq:mu_eq}
\end{equation}

Similarly, the circular dichroism spectra, $\Delta \varepsilon(\omega)  = \eps_L(\omega) - \eps_R(\omega)$, were calculated using the equation of Rosenfeld for a collection of randomly oriented chiral molecules,~\cite{rosenfeld_quantenmechanische_1929}
\begin{equation}
    \Delta \eps(\omega) = \frac{16 \pi \omega }{3 \hbar c} \sum_{a,i} \delta(\omega - \omega_{ai}')\text{Im}[\pmb{\mu}_{ai} \cdot \mathbf{m}_{ia}],
    \label{eq:cdspectra}
\end{equation}
where 
\begin{equation}
    \mathbf{m}_{ai} =  \braket{a|-\frac{1}{2}\mathbf{L}|i}
    \label{eq:m_eq}
\end{equation}    
is the magnetic transition moment from the ground state to exciton $|a\rangle \otimes |i\rangle$ and $\mathbf{L}$ is the orbital angular momentum. The quantity $\text{Im}[\pmb{\mu}_{ai} \cdot \mathbf{m}_{ia}]$ is the rotational strength of the transition.
The Dirac delta function in the above equations is approximated by a Gaussian lineshape broadened by $50$~meV, consistent with the experimental inhomogeneities and finite temperature effects.\cite{lin_theory_2022}

\begin{figure}[t]
    \centering
\includegraphics[width=0.95\linewidth]{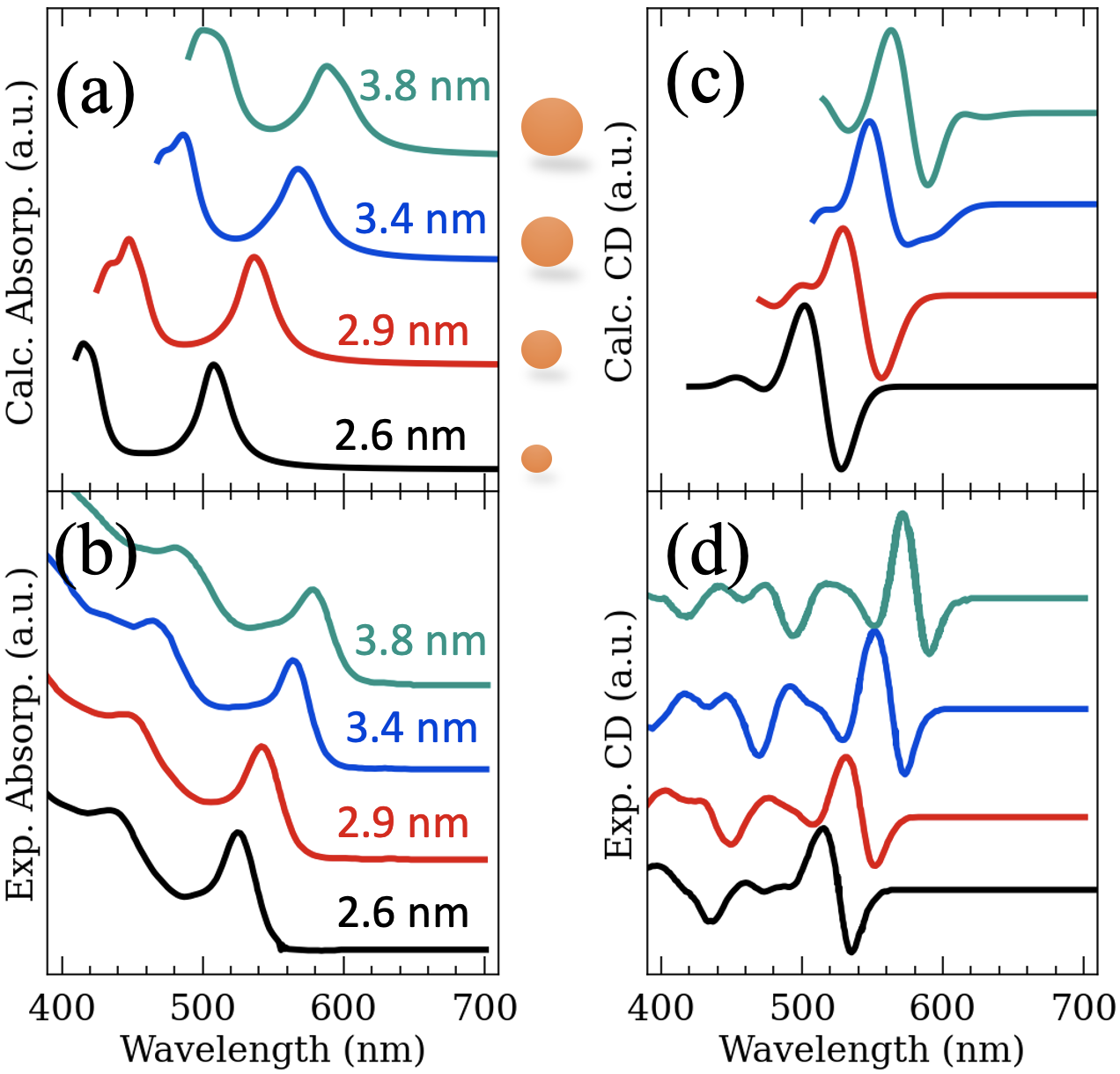}
    \caption{Absorption spectra from theory (a) and experiment (b) as well as CD spectra from theory (c) and experiment (d) for 2.6~nm (black), 2.9~nm (red), 3.4~nm (blue), and 3.8~nm (teal) CdSe QDs. We quantitatively capture the redshift with increasing system size and bisignate lineshape of CD at the low energy excitonic transitions. CD magnitudes have been normalized to appear on the same scale.}
\label{fig:cd_allsizes}
\end{figure}

We apply the described electronic structure model to compute CD spectra for chiral CdSe QDs with diameters of $2.6$, $2.9$, $3.4$, and $3.8$~nm that are comparable to experimental measurements.\cite{ben-moshe_probing_2016, choi_chirality_2016} Fig.~\ref{fig:cd_allsizes} shows a comparison between theoretical and experimental absorption (panels (a), (b)) and CD (panels (c), (d)) spectra. We focus on optical transitions that give rise to the low energy features in the CD spectra. 

As can be seen in Fig.~\ref{fig:cd_allsizes} (panels (a) and (b)), we find very good agreement of the absorption onset comparing the measured and calculated spectra for the range of QD sizes considered. This agreement is expected based on previous validations of the semi-empirical pseudopotential approach,\cite{jasrasaria_simulations_2022}  but does not necessarily extend to calculation of CD response. The agreement in comparison to experiments~\cite{ben-moshe_probing_2016} shown in Fig.~\ref{fig:cd_allsizes} (panels (c) and (d)) suggest that the semi-empirical pseudopotential model can also capture the features at the onset of CD response in QDs, which is not a trivial result. Our model quantitatively recovers the consistent bisignate lineshape observed in CD measurements as well as its redshift with increasing system size.\cite{ben-moshe_probing_2016} This agreement suggests that the CD response is sufficiently described by our ligand potential approach and induced chirality can be understood by an electronic coupling mechanism.

Discrepancies in the size-dependent behavior have been reported in the literature,\cite{ben_moshe_size_2011, tohgha_ligand_2013} and we applied our model to investigate the relationship between QD size and CD strength. Contrary to the increasing absorption trend with size,~\cite{striolo_molecular_2002, jasieniak_re-examination_2009} we find that the CD intensity decreases with increasing QD size.

\begin{figure}[t]
    \centering
    \includegraphics[width=\linewidth]{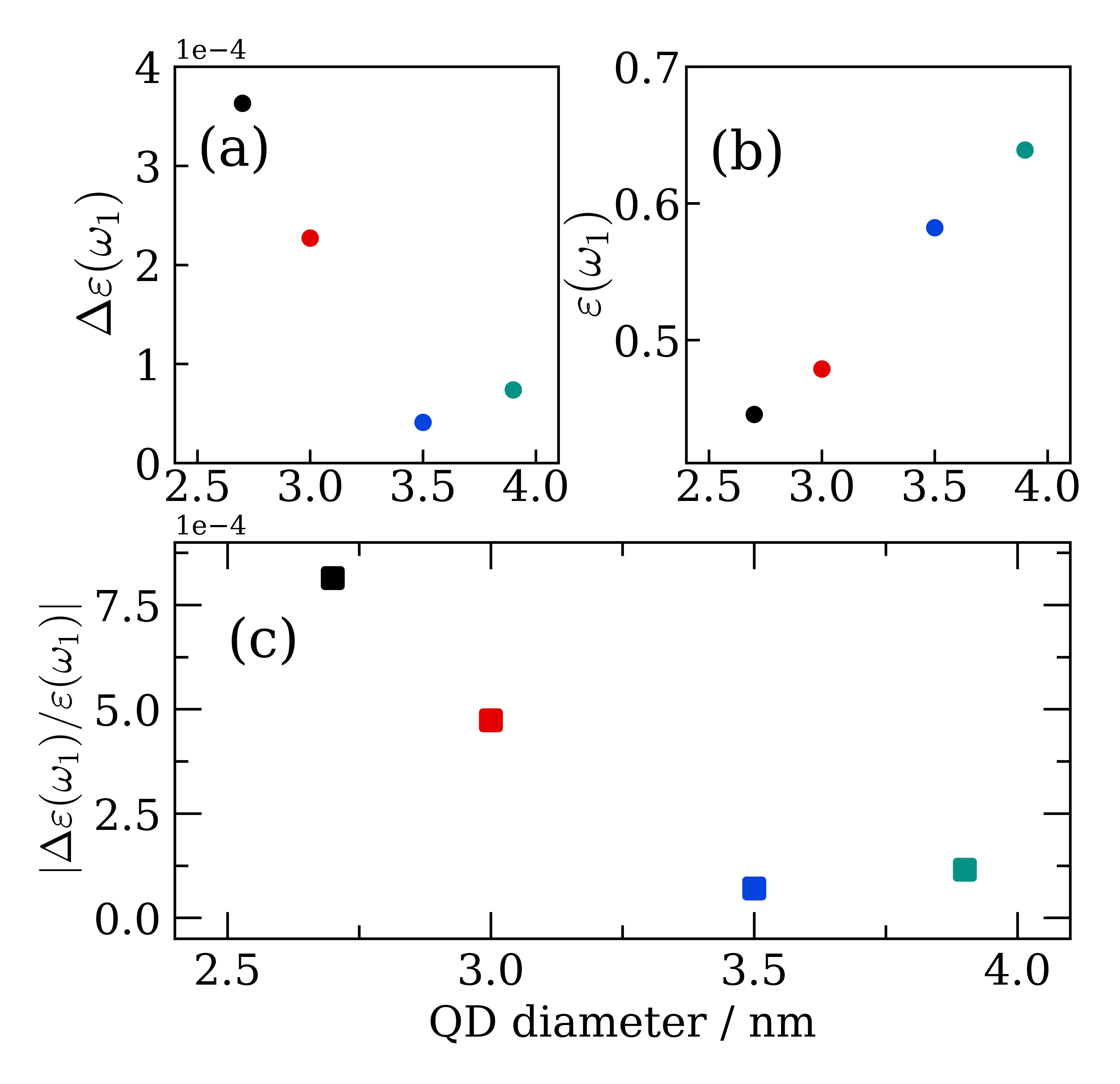}
    \caption{Dissymmetry factors, $g(\omega_1)=\Delta \varepsilon(\omega_{1})/\varepsilon(\omega_1)$, where $\omega_1$ is the frequency of the first CD extrema, plotted against QD diameter. Panel (a) shows the size-dependent trends in CD, while (b) shows absorption. Panel (c) gives the size dependent g-factors. The observed decrease in dissymmetry with size is due to the decrease in CD strength and increase in absorption strength with QD diameter. The magnitude of $g(\omega)$ is arbitrarily controlled by our choice of CLP strength and not associated with any specific passivation ligand.}
    \label{fig:sizedepg}
\end{figure}

The CD response was compared between dot sizes by calculating the dissymmetry, $g(\lambda)$, a measure of optical activity that is scaled by the absorption strength,
\begin{equation}
g(\omega) = \frac{\Delta\varepsilon(\omega)}{\varepsilon(\omega)},
\label{eq:g}
\end{equation}
where $\omega$ is the incident frequency, and the total absorption includes contributions from magnetic dipole transitions. A decrease in $g(\omega)$ could arise due to a decrease in $\Delta \varepsilon(\omega)$, an increase in $\varepsilon(\omega)$, or the simultaneous increase of both with steeper increase in $\varepsilon(\omega)$. The absorption coefficient scales with QD diameter, increasing due to the increase in wavefunction overlap for larger QDs. The differential absorption coefficient, $\Delta \varepsilon(\omega) = \varepsilon_L(\omega) - \varepsilon_R(\omega)$, does not obey a similar trend, but decreases with increasing QD size (Fig.~\ref{fig:sizedepg}). Even though the individual coefficients of absorption for left and right handed light would increase with QD size, their difference depends on the rotatory power of the wavefunctions involved in the electronic transition. A nonzero rotatory strength is induced by the chiral ligands on the surface of the QD. For small QDs, perturbation by the chiral ligand potentials has a large, asymmetric influence on the shape of the QD wavefunctions, leading to large CD magnitude. For large QDs, the chiral ligand potentials on the surface have a relatively weaker influence on the overall asymmetry of the interior QD wavefunctions because the perturbing ligands are small compared to the total wavefunction. This can be interpreted in terms of the surface-to-volume ratios for small and large QDs, where the decreasing surface-to-volume ratio with increasing QD diameter means less influence from the surface ligand perturbation on the total wavefunction dissymmetry.

\label{subsec:bisignate}
A puzzling observation in early chiral QD experiments was the bisignate (derivative) lineshape of their CD spectra,\cite{ben-moshe_probing_2016} very distinct from the CD spectra of molecules, which typically show maxima at the absorption peak.\cite{note-cd}  Equipped with the theoretical tools to investigate CD features from individual excitonic contributions, we first assess the hypothesis of chirality-inducing exciton splitting. It has been proposed that the bisignate lineshape is due to degeneracy lifting of exciton levels under a chiral ligand field, with oppositely polarized magnetic dipole transitions. This phenomenon would lead to neighboring oppositely signed peaks and results in the bisignate shape characteristic of chiral ligand passivated QDs.  

\begin{figure}[ht!]
    \centering
\includegraphics[width=0.95\linewidth]{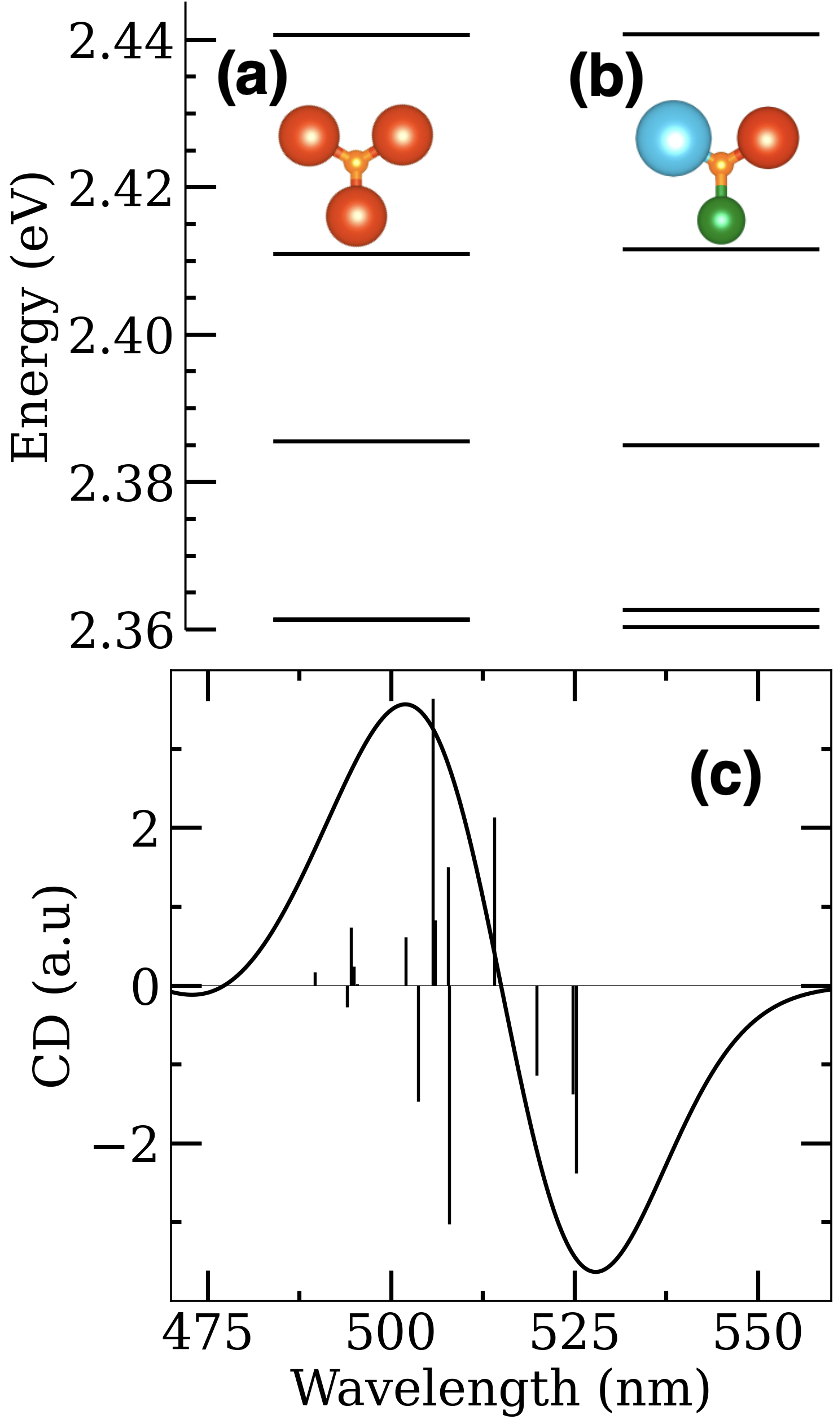}
    \caption{Exciton energy levels for a) an achiral and b) a chiral 2.6 nm CdSe QD. The achiral dot was prepared using the intermediate strength potential at all tetrahedral positions so that the energy scale of the two Hamiltonians were equivalent. The excitonic states are analogous; a small degeneracy breaking of the first two excitonic levels is observed due to symmetry lowering in the chiral structure. However, the energy splitting of the symmetry breaking it not sufficient to explain the bisignate lineshape. Additionally, the rotational strengths of those states are both negative, yielding the same preference of circularly polarized light. c) CD spectra with associated rotational strength sticks for the chiral 2.6 nm QD. }
    \label{fig:exciton_pub}
\end{figure}

The calculated excitonic energies, however, show little evidence for this when comparing achiral and chiral QDs. Fig.~\ref{fig:exciton_pub} shows that the exciton level structure for several low-lying states is minimally affected by the introduction of chiral ligand potentials. A small degeneracy breaking of the first two excitonic levels is observed due to symmetry lowering in the chiral structure. However, the energy splitting is not sufficient to explain the location of the two peaks in the bisignate lineshape. Additionally, the rotational strengths of those states are both negative, yielding the same preference of circularly polarized light. Our model suggests that the observed lineshape emerges from contributions of many transitions of opposing angular momenta (Fig.~\ref{fig:exciton_pub}(c), black line). This behavior emerges from the high density of excitations, differing from the behavior of small clusters and molecules.

\begin{figure}[t]
    \centering
    \includegraphics[width=0.95\linewidth]{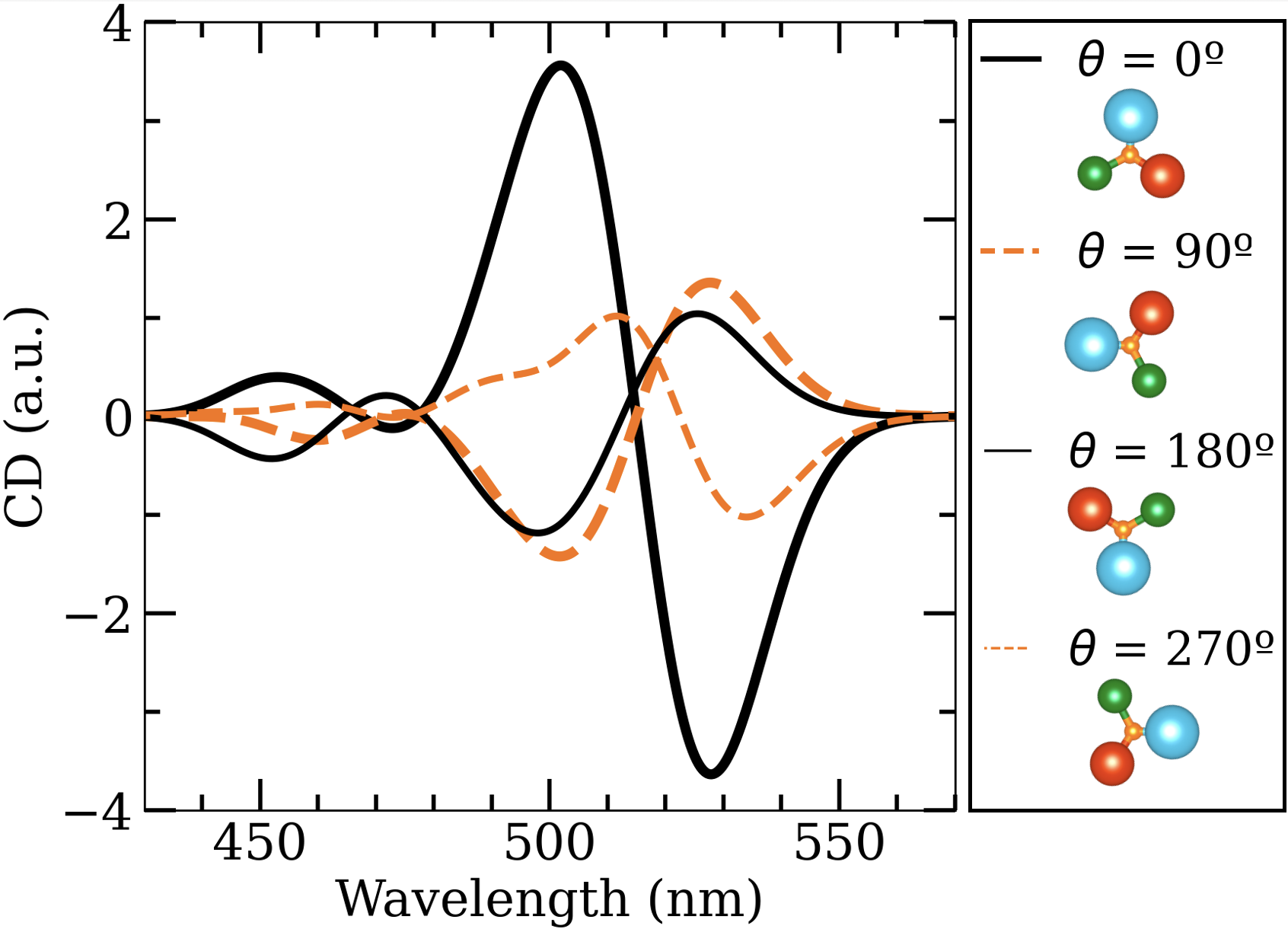}
    \caption{CD spectra of 2.6 nm CdSe QDs with chiral ligand potentials oriented with relative angles of $0^\circ$ (thick black), $90^\circ$ (thin black), $180^\circ$ (thick dotted orange), and $270^\circ$ (thin dotted orange). The spectra reveal strong orientation dependence of the CD signal, with ligand rotations inducing magnitude, sign, and peak wavelength changes. The flipped ligand orientations, $\Delta \theta = 180^\circ$, do not show flipped CD spectra because the asymmetry of chiral ligands enforces a directionality that makes all orientations unique.}
    \label{fig:rot_cd}
\end{figure}

The competing effect of positive and negative contributions means that the wavelength of each derivative peak does not align with the absorption maximum. As further discussed below, modulating the magnitude of the rotational strengths by rotating the chiral ligands shifts the position of the apparent CD extrema, see Fig.~\ref{fig:rot_cd} below for further analysis.

The atomistic control leveraged in our model lends naturally to studying the impacts of the ligand orientation on the CD spectra. To explicitly study the impact of ligand orientation on CD features, we generate configurations where all chiral ligands are oriented with the same angle $\theta$ as defined previously.  We illustrate the orientation dependence for a QD with a diameter of $2.6$~nm with qualitatively similar results for large QDs. 

Fig.~\ref{fig:rot_cd} shows four representative CD spectra with different relative ligand orientations. The lineshapes show a clear dependence on the chiral ligand potential (CLP) angle, with complete inversion of the chiral signal possible through merely a collective ligand rotation. Notice that flipping the chiral ligand from $0^{\circ}$ (thick black) to $180^{\circ}$ (thin black) does not produce the mirror image CD signal. Similarly, the $90^{\circ}$ (thick dotted orange) / $270^{\circ}$ (thin dotted orange) pair does not exhibit mirrored spectra. That is because these complimentary orientations are not mirror symmetric: there are no possible rotations that exactly invert the stereochemistry of a chiral ligand. Chirality enforces asymmetry so that all rotations are unique. 

This directional asymmetry is crucial to understand experiments, where intuition might expect that ensemble distributions of ligand orientations could lead to cancellation of the CD signal. We anticipate that the CD signal will not completely cancel even under isotropic rotations of an ensemble of ligands because the rotational strength depends anisotropically on ligand orientation. This will play a significant role for determining the average CD response of an ensemble of randomly oriented chiral ligand passivated QDs. In addition, we find that the apparent location of each bisignate extrema shifts by up to $10~\mbox{nm}$ in wavelength under ligand rotation.

The sign flips and magnitude changes are understood from the behavior of the underlying transitions. To build up this understanding from a simple conceptual picture, we consider first the rotatory strength of uncorrelated electron-hole pairs. In this limit, a single transition will contribute a rotational strength proportional to $\boldsymbol{\mu}_{ai} \cdot \boldsymbol{m}_{ia} = \mu_{ai} m_{ia} \cos \phi$ as given by Eq.~\eqref{eq:cdspectra}, for an electron in quasi-state $a$ and a hole in quasi-state $i$. The linear dipole transition moment depends on the overlap of an electron in state $\ket{a}$ and a hole in state $\ket{i}$, and the magnetic transition moment depends on the orbital angular momentum component between $\ket{a}$ and $\ket{i}$. The angle $\phi$ is the relative angle between those transition moment vectors. 

We use weak chiral potentials such that the orientation of chiral ligands does not change the magnitude of $\boldsymbol{\mu}_{ai}$ but does lead to changes in the angle, $\phi$. Taking these results together with Eq.(~\ref{eq:cdspectra}) implies that the orientation dependence of the rotational strength is due to changes in $\phi$. In achiral systems, the electric and magnetic dipole vectors are orthogonal, $\phi = \pi/2$, and the rotatory strength vanishes.\cite{tepliakov_optical_2016} But chiral ligands perturb the direction of the magnetic dipole away from orthogonality such that $\phi = \pi/2 \pm \delta \phi$, with $\delta \phi$ a small angle that depends on the orientation of the ligands. Some orientations correspond to positive values of $\delta \phi$ while others to negative values, giving rise to opposite CD response for the same absolute stereochemistry, as depicted in Fig.~\ref{fig:rot_cd}.

This analysis applies to a single quasiparticle transition. To delineate the role of $\phi$ on the observed CD spectra, one has to consider that each optical excitation should be represented by an admixture of quasiparticle transitions and that the overall CD spectrum is a combination of many such excitonic transitions. This makes it difficult to construct an analytic expression for the orientation dependence of the CD spectra, guiding us toward qualitative, intuitive predictions. But two general conclusions can be stated at this point. First, we anticipate that disorder of the ligand orientations, either on a single QD or across an ensemble of dots, will decrease the overall CD signal as positive and negative peaks undergo cancellations. Directly inferred from this is that ordered configurations of chiral QDs will have enhanced CD signals, suggesting that the CD response could report on the degree of ligand shell order.

Recent interest in the optical activity of QDs has raised several questions regarding the CD lineshape and its dependence on size and shape.  To address this, we developed an approach to induce chirality by surface ligands within the atomistic pseudopotential model including electron-hole correlations. We applied the formalism to calculate the size- and orientation-dependent CD spectra for CdSe nanocrystals of varying sizes. We find that the origin of the bisignate lineshape is due to many nondegenerate excitonic transitions of opposing angular momenta that combine to form the characteristic CD features. In addition, we investigated the dependence of the CD response on ligand orientation and found that the relative orientation of chiral ligands modulates the sign, magnitude, and apparent peak locations of the CD spectrum. We conclude that QDs exhibit flexible CD lineshapes based on the precise orientation of passivating ligands. This suggests that chiral ligand tags could enable experimental studies of the order-disorder transition of ligand shells on colloidal semiconductor nanocrystals, lighting the way for chiroptical control.\\

Supporting Information: pseudopotential parameters; GitHub repository with source code, quantum dot configurations, and raw data files; induced CD convergence tests.

\begin{acknowledgement}
This work was supported by the U.S. Department of Energy, Office of Science, Office of Basic Energy Sciences, Materials Sciences and Engineering Division, under Contract No. DEAC02-05-CH11231 within the Fundamentals of Semiconductor Nanowire Program (KCPY23). Computational resources were provided in part by the National Energy Research Scientific Computing Center (NERSC), a U.S. Department of Energy Office of Science User Facility operated under contract no. DEAC02- 05CH11231. We would like to thank Drs. Assaf Ben-Moshe, Ming Chen, John Philbin and Daniel Weinberg for helpful discussions. 
 
\end{acknowledgement}

\bibliography{chiral-qd,misc}

\begin{mcitethebibliography}{61}
\bibitem[not()]{note-cd}
CD spectra of chiral ligand passivated QDs are zero at the absorption maximum and have a bisignate rotatory dispersion reminiscent of the Cotton effect in birefringent molecules.\relax
\mciteBstWouldAddEndPunctfalse
\mciteSetBstMidEndSepPunct{\mcitedefaultmidpunct}
{}{\mcitedefaultseppunct}\relax
\EndOfBibitem
\end{mcitethebibliography}


\end{document}


\maketitle

\section{Pseudopotential Parameters}

We compute the electronic states of CdSe QDs using the semi-empirical pseudopotential approach based on the following form fitted in reciprocal space,\cite{wang_pseudopotential_1996, wang_electronic_1999}
\begin{equation}
    v_\mu(q) = \frac{a_{0,\mu}(q^2 - a_{1,\mu})}{a_{2,\mu}\text{exp}(a_{3,\mu}q^2) - 1}.
\end{equation}
where $\mu$ is the identity of the atom that the pseudopotential describes. The parameters $a_0 - a_3$
were fitted to bulk band structures and effective masses of CdSe obtained by
first principles or experimental measurements.\cite{wang_electronic_1999, cohen_band_1966}
\begin{table}[]
    \centering
    \begin{tabular}{ccccc}
    \hline
     & $a_0$ & $a_1$ & $a_2$ & $a_3$  \\
     \hline \hline
     Cd & -31.4518 & 1.3890 & -0.0502 & 1.6603 \\
     \hline
     Se & 8.4921 & 4.3513 & 1.3600 & 0.3327\\ 
     \hline
    \end{tabular}
    \caption{Pseudopotential parameters for Cd and Se. All parameters are given in atomic units.}
    \label{tab:potparams}
\end{table}

The open valences of surface atoms in the nanocrystal were passivated with ligand potentials designed to push the energies of surface-localized states away from the band edge, the form of which were taken as simple Gaussians.\cite{wang_electronic_1999} The Cd and Se passivation potentials are labeled $P_1$ and $P_2$ respectively. Chirality was induced through the placement of chiral ligand potentials (CLPs) onto the $P_1$ passivation potentials. All $P_1$ ligands were decorated with the chiral arrangements except in cases where high ligand density would cause steric clashing of the CLP lobes. Each CLP is a Gaussian,

\begin{equation}
    v_{\xi}(\mathbf{r}) = \sum_{i=1,2,3} a_{i} \text{exp}\left(- b_i |\mathbf{r - R}_{\xi, i}|^2\right)
\end{equation}
where $\xi$ indexes the CLP attached to a particular $P_1$ passivation ligand, and the paramaters $a$ and $b$ are tabulated below for all ligands.

\begin{table}[]
    \centering
    \begin{tabular}{ccc}
    \hline
     & $a$ & $b$  \\
     \hline \hline
     $P_1$ & 0.64 & 2.229\\
     \hline
     $P_2$ & -0.384 &  2.229 \\ 
     \hline
     $\xi_1$ & 0.05 & 0.16 \\ 
     \hline
     $\xi_2$ & 0.065 & 0.16 \\ 
     \hline
     $\xi_3$ & 0.08 & 0.16 \\ 
     \hline
    \end{tabular}
    \caption{Gaussian parameters for the passivation potentials $P_1$ and $P_2$ and the three CLP lobes. All parameters are given in atomic units.}
    \label{tab:potparams}
\end{table}

\section{Configuration files, chiral QD generation, and electronic structure code}

A series of $2.7 - 3.9$~nm wurtzite CdSe QDs were examined in this study. All configurations, code to generate the chiral QDs, executables to obtain quasiparticle and correlated states, and the generated data can be found on \href{https://github.com/dchabeda/Chiral-QDs.git}{GitHub}.

\section{Convergence Tests}

\subsection{Grid Convergence}

We represent the semi-empirical pseudopotential on a real-space grid. The continuity of the pseudopotential depends strongly on the density of grid points, thus it is necessary to converge each property with respect to the grid density. This poses a particular challenge in this study because the CLPs are rotationally symmetric, which is not commensurate with the rectangular grid. To avoid discontinuities in the potential for rotated CLPs, we made the amplitudes of each CLP small and the width of the Gaussian functions (FWHM) $\approx$ 4.0~Bohr. A grid spacing of 0.6 Bohr is sufficient to smoothly capture the effect on CD of $1^\circ$ ligand rotations. The chosen grid size is also sufficient to converge the excitonic state energies, oscillator strengths, and rotational strengths. An animation of the evolution of the CD spectrum as the ligand rotation angle progresses can be found in the GitHub repository.

\subsection{Chiral Ligand Strength}

We chose the strenght of the CLPs such that they pose a linear response on the QD electronic structure. This ensures that the physical effects captured in the pseudopotential fitting procedure are retained in the calculated chiroptical excitations. We validate our chosen potential strength by computing exciton energies, absorption strengths, and CD strengths for increasing values of the CLP magnitude. Specifically, we vary the parameter $\lambda$ in the expression below,

\begin{equation}
\lambda v_{\xi}(\mathbf{r}) = \lambda \sum_{i=1,2,3} a_{i} \text{exp}\left(- b |\mathbf{r - R}_{\xi, i}|^2\right)
\end{equation}

and plot the resulting observables.

\begin{figure}
    \centering
    \includegraphics[width=\linewidth]{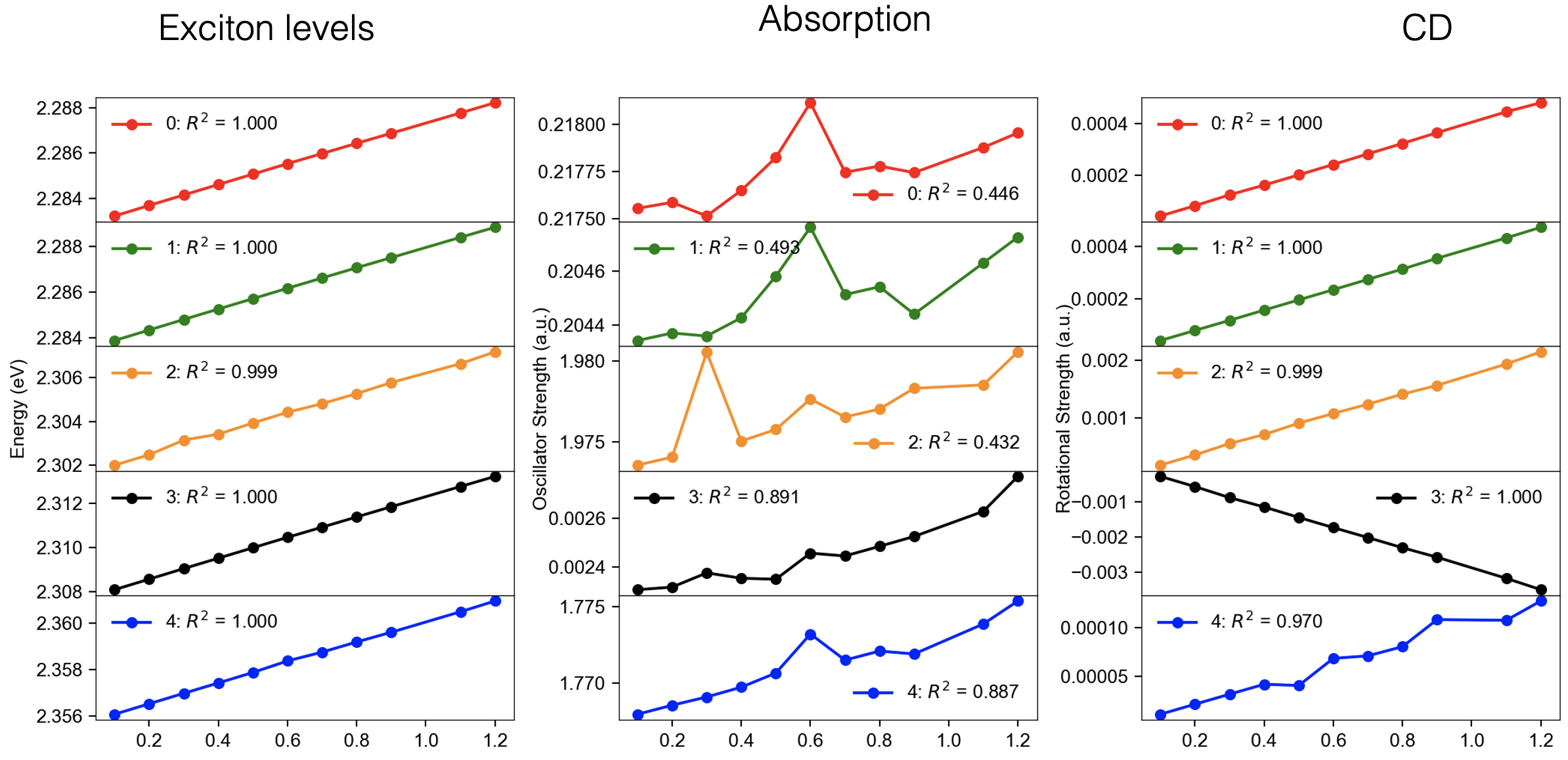}
    \caption{Demonstration that the chosen CLP parameter regime is a small, linear perturbation to the QD properties. Exciton energies, absorption strengths, and CD strengths for the lowest five excitons are shown; the top is the lowest energy and the bottom is the highest energy exciton. The x-axis is the magnitude of, $\lambda$, a tunable prefactor multiplying the CLP Gaussians.}
    \label{fig:linresponse}
\end{figure}

\bibliography{si}